\documentclass[
amsmath,
amssymb,
prd,
twocolumn,
showpacs,
superscriptaddress,
preprintnumbers]{revtex4}
\usepackage{graphicx}
\usepackage{epsfig}

\newcommand{\beq}{\begin{equation}}
\newcommand{\eeq}{\end{equation}}
\newcommand{\be}{\begin{eqnarray}}
\newcommand{\ee}{\end{eqnarray}}

\def\+{\dagger}

\def\simlt{\stackrel{<}{{}_\sim}}
\def\simgt{\stackrel{>}{{}_\sim}}
\def\nn{\nonumber}
\begin{document}
\title{ Cold Dark Matter as Compact Composite Objects}
\author{  Ariel Zhitnitsky}

\affiliation{ Department of Physics and Astronomy, University of British 
Columbia, Vancouver, BC, V6T 1Z1, CANADA}
%\date{July 19.2006}

\begin{abstract}
Dark Matter (DM) being the vital ingredient in the cosmos, still
  remains  a mystery.  Standard  assumption is that the collisionless  cold dark matter (CCDM)
particles are   represented by some  weakly interacting fundamental fields which can not  be 
  associated with any standard quarks or leptons. 
   However, recent  analyses of structure on galactic and sub-galactic scales have suggested  discrepancies and stimulated numerous alternative proposals including, e.g. 
  Self-Interacting dark matter, Self-Annihilating dark matter,
  Decaying dark matter, to name just a few.
  We propose the  alternative   to the standard 
 assumption  about the nature  of DM particles (which
are typically assumed to be   weakly interacting fundamental point -like particles, yet to be discovered).
Our proposal is based on the idea 
that DM  particles are
strongly interacting composite macroscopically large objects which made of
 well known light quarks (or even antiquarks). The required weakness  of the 
   DM particle interactions   
  is guaranteed  by a small geometrical factor
  $\epsilon\sim  \frac{area}{volume}\sim B^{-1/3}\ll 1$ of the composite objects 
  with a large baryon charge $B\gg 1$,
  rather than by a    weak    coupling constant of a new field.
 We argue that the interaction between hadronic 
matter and  composite dark objects 
does not spoil the desired properties of the latter as cold matter.
We also argue that  such a scenario does not contradict to the current observational 
data. Rather, it has  natural explanations of  many observed data, such as 
 $\Omega_{DM}/  \Omega_{B}\sim 1$ or $511~ KeV$ line from the bulge of our galaxy.
We also  suggest that composite dark matter  may 
modify the dynamics of structure formation in the central overdense regions 
of galaxies.
We also present a number of other cosmological/astrophysical observations which   indirectly  support  the novel concept   of DM nature.

\end{abstract}
\pacs{98.80.Cq, 95.30.Cq, 95.35.+d, 12.38.-t}
\maketitle

\section{Introduction}
  
Observational precission data gathered during the last fifteen years 
have guided the development of the so called concordance cosmological 
model $\Lambda$CDM \cite{Bahcall:1999xn} of a flat universe, $\Omega 
\simeq 1$, wherein the visible hadronic matter represents only $\Omega_B 
\simeq 0.04$ a tiny fraction of the total energy density. Most of the 
matter component of the universe is thought to be stored in some unknown 
kind of cold dark matter, $\Omega_{DM} \simeq 0.24$. The largest  
contribution $\Omega_{\Lambda} \simgt 0.70$ to the total density  
  is  cosmological dark energy with negative pressure, another mystery which 
  will not be discussed here.
 
 There is a fundamental difference between dark matter
  and ordinary matter (aside from the trivial difference
 dark vs.  visible). Indeed, 
 DM played a crucial role in the formation of the present  structure in the universe.  Without dark matter, the universe would have remained too  uniform to form the galaxies.  
 Ordinary matter could not produce fluctuations to create any significant  structures   because it remains tightly coupled to radiation, preventing it from clustering, until  recent epochs.   
On the other hand, dark matter, which is not coupled to photons, would permit tiny  fluctuations  
 to grow for a long, long time  before the ordinary matter decoupled from radiation.  Then, the ordinary matter would  be rapidly drawn to the dense clumps of dark matter and form the observed structure.   
The required material is called 
the  Cold Dark Matter (CDM), and  the obvious candidates are weakly interacting 
particles of any sort which are long-lived, cold and collisionless. 
While this model works very well on large scales, a number of discrepancies have arisen
between numerical simulations and observations on subgalactic scales, see 
 \cite{Spergel:1999mh},  \cite{Dave:2000ar},  \cite{Ostriker:2003qj}, \cite{Primack:2002th}, 
   \cite{Bertone:2004pz} and references therein.
Such discrepancies have stimulated numerous alternative proposals including, e.g. 
  Self-Interacting dark matter, Self-Annihilating dark matter,
  Decaying dark matter, and many others,
  see   \cite{Ostriker:2003qj}  and references therein. 
  There are  many other cosmological/astrophysical observations
   which  apparently  also  suggest  that  
 the standard assumption (that the  dark matter made of   
absolutely stable and  ``practically  non-interacting" fundamental particles)
is oversimplified. Some of the observations    that may be in conflict with the standard
 viewpoint are:
 
 $\bullet$ The density profile is too cuspy,  \cite{Ostriker:2003qj}, \cite{Primack:2002th}, 
   \cite{Bertone:2004pz}.  The disagreement of the observations with high resolution
   simulations is alleviated   with time, but some questions still remain 
     \cite{Primack:2002th},   \cite{Bertone:2004pz}.
 
  $\bullet$ The number of dwarf galaxies in the Local group is smaller  than predicted by CCDM simulations,    \cite{Ostriker:2003qj}, \cite{Primack:2002th}, 
   \cite{Bertone:2004pz}. This  problem is also becoming  less dramatic  with time
     \cite{Primack:2002th},   \cite{Bertone:2004pz}.
   
   $\bullet$ CCDM simulations produce galaxy disks that are too small and have too little angular momentum,   \cite{Ostriker:2003qj}, \cite{Primack:2002th}, 
   \cite{Bertone:2004pz};

$\bullet$  There is a close relation between rotation curve shape and light distribution.
This implies that there is a close coupling between luminous and dark matter 
which is difficult to interpret, see e.g. \cite{Sancisi:2003xt};

$\bullet$ There is a correlation in early -type galaxies 
supporting the hypothesis that there is a connection between the DM content and the 
evolution of the baryonic component in such systems, see e.g.\cite{Napolitano:2003xm};

$\bullet$ The order parameter (either the central density or the core radius) correlates with the stellar mass in spirals\cite{Salucci:2005mi}.
This suggests the existence of a well-defined scale length in dark matter haloes, linked to the luminous matter, which is totally unexpected in the framework of CDM theory, but 
 could be a natural consequence of   DM and 
baryon interaction.

$\bullet$  There is a mysterious correlation between visible and DM distributions 
on $\log -\log$ scale, which is very difficult to explain within the standard 
CCDM model \cite{Mishchenko:2003in};

$\bullet$ A recent analysis of the CHANDRA image of the galactic center
finds that   the intensity of the diffuse  $X$ ray emission 
significantly exceeds the predictions of a model  which includes known Galactic sources
\cite{Muno:2004bs}. The spectrum is consistent with hot $8 ~ KeV$  spatially  uniform plasma.
The hard X-rays are unlikely to result from undetected point sources, because no known population of stellar objects is numerous enough to account for the observed surface brightness.

We shall argue below  that the  observed excess  of the diffuse  $X$ ray emission  
may be originated from  
 DM component with non- negligible interactions  with baryons and photons;

$\bullet$ A recent analysis of the EGRET data finds the intensity of
 the $GeV$  $\gamma$ ray component  
significantly exceeds the predictions of a model  which includes known Galactic sources.
More than that, the excess in different sky directions has identical  energy spectrum, see 
e.g. ref.\cite{deBoer:2004ab} for a nice 
review of data. This observation strongly suggests that   the  excess  may be originated from  
 DM component with non- negligible interactions with itself or/and with baryons\footnote{
 One should remark here that the interpretation of the same data could be very different:
 the agreement with EGRET data can be achieved by  
 assuming a specific variation of cosmic rays intensity  in space and time. It also requires some
 adjustments of the spectrum of cosmic rays   at low energies. This subject is still matter
 of debates and we refer to the original literature for details \cite{Strong:2004de}.  }.
 
 $\bullet$ The soft gamma-ray spectrum in $1 -20 ~MeV $ region cannot fully be attributed to either Active Galactic Nuclei or Type Ia supernovae or a combination of the two \cite{MeV}; thus,  some sort of 
 ``interacting"   dark matter particles may be required for a  possible explanation for soft gamma-ray spectrum  in $MeV$ region;
 
  $\bullet$ Related, but still, a separate  issue is the observation  of
 $511$~KeV $\gamma$-ray line from the bulge of the Galaxy 
with spherically symmetric distribution \cite{Knodlseder:2003sv}.  The intensity 
and some features   of the flux are such that it is quite difficult to
explain by known  astrophysical processes.  This
observation also strongly suggests that   the  excess  may be related to   
some sort of  DM component with non- negligible interactions with photons.

This list of questions above
is obviously very  far to be   complete. The list of references \cite{Spergel:1999mh}
- \cite{Knodlseder:2003sv} (where these questions 
have been discussed) 
even less complete.  However, the main point we want to make here is as follows:
it appears that the DM and ordinary baryons somehow ``know" about each other,
(beyond the trivial gravitational interaction).
Each piece of evidence taken separately is perhaps not convincing enough to 
  abandon the idea that DM is collisionless, non-interacting and  absolutely stable
  weakly interacting massive particle.  Nevertheless, it is very likely 
  that some of the problems (mentioned and not mentioned above) 
  persist. In this case it would be an indication that DM is not as trivial  substance as it   thought to be.
 In fact, motivated by first three items above,  it has been suggested recently,
 \cite{Spergel:1999mh} that DM is  actually   Self-Interacting dark matter
 (SIDM) with strength which encompassed the range 
 \beq
 \label{s}
 s=\frac{\sigma_{DD}}{M}\simeq (8 \cdot 10^{-25}- 1\cdot 10^{-23})\frac{cm^2}{GeV} ,
  \eeq  
  see also earlier work on the subject \cite{Starkman:1990nj}.
  This   scale is so similar to the  typical cross section  for ordinary hadrons 
  at low energies, that it  has been even assumed \cite{Wandelt:2000ad} that the DM is composed 
  of exotic hadrons such that interaction between DM particles and ordinary hadrons is  
   the same order of magnitude as given by  eq.(\ref{s}). Many models with such strong interaction of  DM 
   with ordinary hadrons are probably already ruled out e.g. from analysis of cosmic rays -DM interactions
   \cite{Cyburt:2002uw}, or from some other  constraints.
   
    However, a general idea that DM could be an object  strongly interacting   with ordinary baryons
   ( in view of many hints  coming from  very different unrelated observations,  see  some  highlights  above)     still remains to be a very attractive idea.   
   
  In fact, it was  recently suggested  a natural reason 
  why the dark matter objects might be closely related to the ordinary baryons \cite{Zhitnitsky:2002qa}, \cite{Oaknin:2003uv}.
 Our original  argument   suggesting the necessity of such kind of connection was based on 
    the observation  that $\Omega_B\sim \Omega_{DM}$. 
    Indeed, these two contributions to $\Omega$ could be in general 
   very different   because (according to the canonical view) they are originated from fundamentally 
   different physics at very different cosmological epoch. Therefore, the observed relation
  $\Omega_B\sim \Omega_{DM}$ between the two very different contributions
to $\Omega$ is extremely difficult to explain in models that invoke a   DM candidate
not related to the ordinary quark/baryon degrees of freedom. 

   We shall see in what follows, that  a   resolution 
of the puzzle $\Omega_B\sim \Omega_{DM}$ within our framework might be linked to a number of 
other problems highlighted above. We are not claiming, of course, to have these  problems 
 solved in our framework. 
   Rather, we want to present some arguments
 suggesting that many apparently    unrelated problems
  might be in fact closely related.   
  
  The idea is that the  {\it dark matter } consists of very dense 
(few times the nuclear density) macroscopic 
droplets of {\it ordinary  light quarks }( or/and  antiquarks) 
 \cite{Zhitnitsky:2002qa}, \cite{Oaknin:2003uv} which however are formed not  in 
ordinary hadronic phase, but rather in color superconducting phase, similar 
to the Witten's strangelets \cite{Witten:1984rs} with mass $M\sim  B m_p  $,
where $m_p$ is proton mass and $B$ is the baryon charge of a droplet. 
See also \cite{Froggatt:2005fk},\cite{Farrar:2005zd},\cite{Dolgov}
 where   different 
  mechanisms were suggested 
with potential to explain the observed ratio,
 $\Omega_B\sim \Omega_{DM}$. 
 See also \cite{Khlopov:2005ew} where composite dark matter was considered, but in 
 a very different context. 
 
 Therefore, while 
the massive droplets carry a large baryon charge $B\gg 1$, they do not contribute
to  $\Omega_B$, but rather, they    do contribute to the 
``non-baryonic" cold dark matter $\Omega_{DM}$ of the universe
 \cite{Zhitnitsky:2002qa}, \cite{Oaknin:2003uv}, thus making desirable correlation
 between DM and baryons as highlighted above. 
 
 We should make remark here from the very beginning: 
 while the stability of these objects can be analyzed in relatively simple way
 with very specific prediction for    a given model, 
 the estimation of the  probability of formation of
 such objects is much more difficult task. 
In particular, the central value for the baryon charge $B$
 for  the model suggested in \cite{Zhitnitsky:2002qa} is  $B\sim 10^{33} $ 
 (it is interesting to note that this value corresponds to $M\sim 10^{33} GeV\sim 10^6$ kg,
 which is quite  close to the existent 
 constraint obtained from analysis of seismic events\cite{Herrin:2005kb}.) 
 
 Therefore, 
  we are not attempting to estimate abundance of CCOs in the present work, rather we 
  take  an ``observational" attitude.  Let us assume that such
 droplets indeed are formed/survived  during the QCD phase transition. 
 What would be the observational consequences of this ``historical event"?
  \footnote{ The corresponding analysis
on formation  of the  macroscopically  large  objects  is expected to be very complicated problem.
 In particular, 
it  would require knowledge of the non equilibrium  dynamics of the QCD phase transition 
 at nonzero baryon density $\mu\neq  0$  and nonzero value of  parameter $\theta\sim 1$
which  is essential element for the mechanism to be operative as discussed in \cite{Zhitnitsky:2002qa}.
 Unfortunately we have very limited knowledge about the  QCD phase diagram  
  when these parameters are non-zero. The   first steps
 in the study of the QCD phase diagram at  $\mu\neq  0$ and  $\theta\sim 1$
  (with motivation from cosmology)
 have been undertaken only  very recently\cite{Metlitski:2005db}.  
   Nevertheless, few arguments (supporting the idea that   indeed   a 
   separation of baryon and anti baryon charges during the QCD phase
  transition at  $\mu\neq  0$ and  $\theta\sim 1$ is possible if some conditions are met)
  have been presented  in ref. \cite{Oaknin:2003uv}. 
  The most important lesson from this study was that if small droplets are formed (as a result of fluctuations),   they start to grow due to the differences of reflection and transmission coefficients for quarks and antiquarks at $\theta\neq 0$: CCOs made of matter prefer to capture quarks/baryons while
  CCOs made of antimatter prefer to capture antiquarks/antibaryons.
   The difference in coefficients is order of one, therefore the growth of CCOs is extremely
 efficient. However, even it is $ 100\%$ effecient, 
 some baryons along with  CCOs would still remain in our universe 
 because of the large CP violation at that time
 (equal number of CCOs made of matter and antimatter when no baryons left would
 correspond to the exact CP symmetric case). 
   This process of formation continues until  temperature becomes sufficiently small ($T\sim 40~ $MeV)
   when gas of particles becomes sufficiently dilute, and their energies
   are relatively low to penetrate into the droplets. It is expected that the  difference in numbers   
   of CCOs made of quarks and antiquarks will be order of one due to the strong CP violation
   at that time.  This  is  exactly
   the main reason  why $\Omega_{DM}\sim \Omega_B$,
   see eq. (\ref{ratio}). Precise calculation of $T\sim 40~ $MeV represents a  very difficult problem
   of  the non equilibrium QCD.
   However our estimates presented in ref. \cite{Oaknin:2003uv} strongly suggest
   that it falls into the appropriate energy scale.
   As we said in the text, 
 in the present work we just take a simple  ``observational attitude" formulated 
 as follows: let us assume that such
 droplets indeed can be formed during the QCD phase transition. What would be the observational consequences of this ``historical event"? }.  

 Our presentation is as follows. In section II we argue that 
 such a scenario does not contradict the current observations.
  In sections III and IV we put forward this idea, and argue that in fact many
  puzzles formulated in Introduction can be related to each other
  within our framework when ``Nonbaryonic" dark matter  is actually made of strongly 
  interacting {\it ordinary  light quarks} hidden in a  {\it compact composite object (CCO)}.

  \section{``Nonbaryonic" dark matter as compact composite object (CCO)}

We should emphasize from the very beginning that while  our estimates below
are based on  a specific model \cite{Zhitnitsky:2002qa}
 for the composite DM,  the   main concept (and phenomenological consequences) 
 have  much more generic  applicability. 

\subsection{ Baryonic CCO --no contradiction with BBN}
 The main idea is that
the  baryon charge of massive composite droplets  does not change the 
nucleosynthesis calculations because in the color superconducting   phase it 
is not available for nuclearsynthesis 
when the baryon charge is locked in the coherent superposition
of Cooper pairs (with a typical gap $\Delta \simeq 100 ~
MeV \gg T_{BBN}\simeq 1 ~MeV  $). Therefore, while 
the massive droplets carry a large baryon charge, they do not contribute
to  $\Omega_B$, but rather, they      contribute to the 
``non-baryonic" cold dark matter $\Omega_{DM}$ of the universe, see 
\cite{Zhitnitsky:2002qa}, \cite{Oaknin:2003uv}. In this sense there is a fundamental
difference between CCOs and ordinary compact stars  (apart from the  differences in sizes
and formation history):
the quarks forming the compact stars did participate  in BBN, and therefore they 
contribute to $\Omega_B$, while the same quarks forming CCOs did not participate  in BBN
as explained above, and therefore they 
contribute to $\Omega_{DM}$.
 
  The compact   composite objects  can be made from antimatter as well, not necessary 
from matter. Total contribution from these compact   composite objects made of antimatter naturally has
 the same order of magnitude as $\Omega_{DM}$.
Still, it would not   contradict to  BBN due to the same reason: at $T\simeq 1 MeV$ the baryon charge 
from CCO is not available to participate in nuclear synthesis. 
Therefore, only baryon charge in hadronic phase can participate in BBN.
 The  baryon (antibaryon ) charge hidden 
in CCO remains unavailable for BBN and serves as DM.
\subsection{ Antimatter in the form of CCO. \\No contradictions with observations}

It is important to remark here that bounds that tightly constraint the 
presence of significant amount of antimatter in   the universe 
 are mainly derived from the phenomenological 
signatures of electromagnetic matter-antimatter annihilation processes 
\cite{Dolgov:1991fr}. These bounds do not strictly apply to the presence 
of antimatter stored in CCO as such 
kind of objects do not easily annihilate. Or to say it more precisely,
the rate of annihilation is highly suppressed due to the very small volume occupied
by the objects.

Our scenario is based on the idea that while the universe is globally symmetric,
the antibaryon charge can be stored in 
chunks of CCO antimatter. In different words, 
the baryon asymmetry of the universe may not necessarily be expressed as 
a net baryon number if the anti-baryon charge 
is accumulated in form of CCO, rather than in form of 
free anti-baryons in hadronic phase. 

Such a picture does not contradict the observations.
Indeed, we can estimate the total number of collisions between ordinary 
hadrons and CCOs in a Hubble time.  
The number  density of hadrons in the ordinary phase is, on average, 
$n_B\sim\frac{0.15\rho_{DM}}{1 GeV}$. Thus, the number of collisions per unit time 
in presence of a single CCO is given by
\beq
\label{insert_1}
\frac{d{\widetilde W}}{dt}=4\pi R^2n_{B}v  \simeq 
 4\pi R^2\frac{0.15\rho_{DM}}{1 GeV}v,
\eeq
where $R\sim M^{1/3}\sim B^{1/3}$ is a typical size of CCOs and $v/c\sim 10^{-3}$ is typical velocity
of visible particles.
Even if the annihilation is 100\% efficient,  the total (anti) baryon charge  $\Delta B$  from anti -CCO
which will be  destroyed by such annihilations during a Hubble time does not exceed  
\beq
\label{insert_2}
\Delta B \simeq \frac{d{\widetilde W}}{dt} \cdot H^{-1} \leq 0.1 B^{2/3},
\eeq
per CCO with charge $B$. This represents an exceedingly small
part of the CCO, $\Delta B/B\sim 0.1 B^{-1/3}$ for sufficiently large  $ B$. 
The probability that CCO will collide  another anti CCO during Hubble time is even smaller, 
\beq
\label{insert_3}
\frac{d{\widetilde W}}{dt}\cdot H^{-1}   
 \leq  \frac{0.1}{B^{1/3}}.
\eeq
 If one uses already existing constraint on such kind of objects,  $B \geq 10^{18}$ \cite{Zhitnitsky:2002qa},  or even, $B \geq 10^{30}$ \cite{Herrin:2005kb}, 
one concludes that there is no obvious 
contradiction of the suggested scenario with present observations\cite{Oaknin:2003uv}--
DM in form of CCO and anti CCO  lives much longer than the Hubble time, and can not be easily
destroyed by visible matter\footnote{An obvious question is: what happens when a CCO
hits the Earth. By obvious reasons this question was addressed earlier, in the   first original paper
on the subject \cite{Zhitnitsky:2002qa}. The only comment we would like to make here is as follows.
Due to the gap in color superconducting phases the vast majority of slow non relativistic particles will be 
reflected when they hit CCO. It is equally true for matter as well as for antimatter. Therefore, 
phenomenological consequences of such an event would be very similar to what have been discussed
previously for the Witten's  droplets.  It is expected that such an event would produce a line like
seismic event in contrast with conventional point like events due to the  earthquakes, see
  ref. \cite{Herrin:2005kb} for the detail discussions, latest constraints and earlier references.}.
 
In fact, one can argue that the observed excess of $\gamma$ ray flux 
in $MeV$  and $ GeV$ bands
 might be naturally explained
from the rare events of  annihilation as estimated in eq. (\ref{insert_1}), see next subsection. Also, 
 the observed cosmological ratio between the energy densities
of dark and baryonic matter, $\Omega_{DM}\sim\Omega_{B}$ within an 
order of  magnitude, finds its natural explanation in this scenario: both  
contributions to $\Omega$ originated from the same physics at the same 
instant during the QCD phase transition. 

Indeed, within our framework 
the total baryon number is conserved and,  
therefore,  the net number density of   CCO droplets
should be
\begin{equation}
\label{1}
{\widetilde n}_{\bar{B}} - {\widetilde n}_{B} = \frac{1}{B}( n_B- 
n_{\bar{B}}) \simeq \frac{1}{B} n_B,
\end{equation}
where we introduce notation  ${\widetilde n_B}$ (${\widetilde n}_{\bar{B}}$) 
 describing the number density
of dark matter baryonic (antibaryonic) CCOs  
 which carry the baryon charge in a hidden 
form   rather than in form of free 
baryons. 
Let then consider the ratio of $dark ~matter~ number~ density 
\equiv {\widetilde n}_B + {\widetilde n}_{\bar B}$ to $baryon ~number~ 
density \equiv n_B$. By definition,
\begin{equation}
\label{r1}
\left(\frac{dark ~matter~ number~ density}{baryon~ number~ density}\right)
\simeq \frac{m_N\Omega_{DM}}{M_{DM}\Omega_B }.
\end{equation}
The $dark ~matter~ number~ density$ could be naturally estimated, without 
any {\it fine-tuning}, to be 
\begin{equation}
\label{2}
{\widetilde n}_{\bar{B}} + {\widetilde n}_{B} =C ({\widetilde n}_{\bar 
B} - {\widetilde n}_B),
\end{equation}
where $C$ is some numerical factor $ \simgt 1$, if the excess
$({\widetilde n}_{\bar B} - {\widetilde n}_B)$ is of the same 
order as the number densities ${\widetilde n}_{\bar B}$ and  ${\widetilde 
n}_B$. In fact, the excess $({\widetilde n}_{\bar B} - {\widetilde n}_B)$ 
is indeed expected to be of order ${\widetilde n}_{\bar B}, ~{\widetilde 
n}_B$ if the universe  is largely C and CP asymmetric at the onset 
of formation of the condensed balls. Then, the l.h.s. of the ratio 
(\ref{r1}), can be estimated to be 
\begin{equation}
\label{3}
({\widetilde n}_{\bar B} + {\widetilde n}_B)/n_B 
=C ({\widetilde n}_{\bar B} - {\widetilde n}_B)/n_B  \simeq \frac{C}{B}
\end{equation}
according to eqs. (\ref{1},\ref{2}). 
Consequently, from (\ref{r1},\ref{3}) we obtain 
\begin{equation}
\label{ratio}
\Omega_{DM}/\Omega_B \simeq (C/B) \cdot (M_{DM}/m_N).
\end{equation}
Now, if one demands $\label{input} B \simeq (M_{DM}/m_N)$, which 
is a condition for the  stability of the droplets \cite{Zhitnitsky:2002qa},  one can immediately
derive $\Omega_{DM} / \Omega_B \simeq  C \simgt 1$. The point we want to 
make is: our assumption that the dark matter is originated at the QCD 
scale from ordinary quarks fits very nicely with $\Omega_{DM} / \Omega_B 
\simgt 1$ within the order of magnitude, provided that separation of 
baryon charges is also originated at the same QCD scale. Generally, the 
relation $\Omega_B \simlt \Omega_{DM}$, within one order of magnitude, 
between the two different contributions to $\Omega$ is extremely  difficult to 
explain in models that invoke a dark matter candidate not related to the 
ordinary quark/baryon degrees of freedom. 
The baryon to entropy ratio $n_{B}/n_{\gamma}\sim 10^{-10}$ would also
be a natural outcome in this scenario. 
We refer to the original paper  \cite{Oaknin:2003uv}
for the details.  
\subsection{ Annihilating  CCOs. No contradictions \\with 
$\gamma$ ray flux observations}
The DM in form of CCO, in some sense,  has some features
of  annihilating DM\cite{Kaplinghat:2000vt} as baryon charge from visible matter and antibaryon 
charge from anti CCOs do annihilate as discussed above. 
Naively, one could think that   large
amount of antimatter  (order of $\Omega_{DM}$)  is  already in severe contradiction
with $\gamma$ ray flux observations.  Indeed, in order to avoid the contradictions with  $\gamma$ ray flux observations, 
the authors of the annihilating DM proposal \cite{Kaplinghat:2000vt} have assumed that annihilation 
products must not include photons. In our scenario when
 DM is represented by ordinary quarks/antiquarks we do not have a luxury  to make such kind
 of  assumptions  because   ordinary quarks and anti quarks   do annihilate and do produce photons.
 However, as we shall see, when DM is locked in CCOs   still  there is no contradiction with 
$\gamma$ ray flux observations. In  fact, such annihilation might be a natural solution  of
a long standing  problem on observed $\gamma$ ray excess in $MeV$  and $ GeV$ bands.
By definition, the flux is defined as 
\be
\label{flux}
\Phi = \int ds \int_{\Delta \Omega} d\Omega\frac{dW}{dVdt} (r),
\ee
 where $\frac{dW}{dVdt}(r) $ is  the probability of the annihilation  event per unit volume per unit time
 at point $r$ measured from the center of the galaxy,
  $ \Delta \Omega $ is the solid angle observed, and  the integral  $\int ds$ is performed over the line of sight of the observation.
The   probability of  annihilation can be estimated as in eq.(\ref{insert_1}), 
 \be
\label{p}
\frac{dW}{dVdt}(r) \simeq 
4\pi R^2 \cdot v \cdot n_{B}(r )\cdot n_{DM}(r)\simeq \nn \\
   {4\pi R^2}   \cdot v  \cdot (\frac{\rho_{B}(r)}{1 GeV}) \cdot  (\frac{\rho_{DM}(r)}{B\cdot 1 GeV})\sim B^{-1/3},
\ee
where we assume that annihilation is $100 \%$ efficient such that all baryons hitting the CCOs
will annihilate. One can check that estimate for the flux (\ref{flux})  with $ \frac{dW}{dVdt}(r)$
given by eq.  (\ref{p})    is  not  in contradiction with  observations 
for sufficiently large $B$\footnote{Precise
constraint on $B$ depends on  specific model for   $\rho_{DM}(r)$  and $\rho_{B}(r) $.}.
 This is the  main massage of this subsection.
 
Rather than making constraints on $B$ it is very tempting to assume that the excess of $511 KeV$ photons \cite{Knodlseder:2003sv} 
 as well as $\gamma$ -excess in $MeV$   \cite{MeV} and $GeV$ bands \cite{deBoer:2004ab}
 as highlighted in Introduction, can be  
 explained precisely by this annihilation with dark matter in form of CCOs.
 In fact,   all features of $511 KeV$ line from the bulge of our galaxy 
 (including the width, spectrum  and intensity) can be naturally explained  
 by using eq. (\ref{p}) and accepting the standard distributions for the dark and visible matter
 when $B\sim 10^{33}$ \cite{Oaknin:2004mn}. Once the general normalization is fixed 
 from  the observation of $511 KeV$ line, one can unambiguously predict the flux   integrated over photon's spectrum  in  $MeV$ region originated from annihilation of visible matter 
 with dark matter in form of CCOs\cite{arz}.
   Corresponding calculations are beyond the scope of the present work;
  however,  the obvious consequence of the scenario is 
   that  the flux  in $MeV$ range and  $511 KeV$ line
 must be strongly correlated    in all sky directions. 
 Unfortunately, available data (see  \cite{MeV} and references 
therein) are not sufficient  to make a positive statement on this.
However, what is known is definitely consistent  with this prediction.
 We   also point out that  
  $e^+e^-$ annihilation with a single bright $511 KeV$ line  
   should be accompanied by the wide (70 MeV -1 GeV) 
  $\gamma$  spectral density due to the annihilation
  of  baryon from visible matter with   anti baryonic charge from dark matter in form of CCOs.
   These very different spectra in different frequency regions must be related to each other    due to their  common origin.  
  Corresponding calculations are beyond the scope of the present work;
  however, a very simplified estimate of the corresponding flux 
 can be obtained by replacing electron velocity $v$
  in formula (\ref{p}) by a proton velocity 
  $v_p/v\sim \sqrt{m_e/m_p}\sim 2 \cdot 10^{-2}$\cite{Oaknin:2004mn}.
     This  corresponds to the assumption
 of the thermal equilibrium between electrons and protons 
 in the ionized region  in the bulge of the galaxy.
  Estimated in such a way flux
 is consistent with observations, where some access of $\gamma$ rays  indeed has been observed by EGRET, see  \cite{deBoer:2004ab} and references therein. 
\subsection{ Photon - CCO decoupling. No contradictions with 
structure formation constraints}
As we discussed in previous subsections,
 the interaction between dark matter in form of CCOs and photons 
is quite strong. Therefore, a natural question arises whether such an interaction does not spoil
the main feature of the DM, and whether 
 the compact composite objects of condensed quark matter do indeed qualify as 
candidates for the role of cold dark matter of the universe at the time 
$T_{eq} \simlt 1$~eV when large scale structures develops.   
  Such interaction, in principle,  could spoil the desired non-thermal 
distribution of dark composite objects. However, as we shall estimate below,
sufficiently large CCOs  do indeed qualify as  cold dark matter candidates.

First, we recall why ordinary matter can not play a crucial role in structure formation. 
This is due to the fact that the baryons are tightly coupled to the photons until decoupling time, 
$z_{Dec}\simeq 1100, ~ T_{Dec}\simeq 0.26 eV$. This tight  coupling provides the
baryonic fluid with a pressure which  prevents the
small perturbations to grow due to the force of gravity.

Indeed, at $t\ll t_{Dec}$ photons and baryons are very tightly coupled due to Thomson scattering.
The mean free time
\beq 
\label{Thomson}
t_{Th} =\frac{1}{x_en_B\sigma_{Th}c}\simeq 6\cdot 10^{7} s \left(\frac{T}{1 eV}\right)^{-3}\left(x_e\Omega_Bh^2\right)^{-1}
\eeq
with $x_e$ being the fraction of charged particles, 
and $\sigma_{Th} = \frac{8\pi}{3}(\frac{e^2}{m})^2$ 
being the Thomson scattering cross section, should be compared with Hubble time
\beq
\label{Hubble}
H^{-1}\simeq 1.13 \cdot 10^{12}s \left(\frac{T}{1 eV}\right)^{-3/2}\left(\Omega h^2\right)^{-1/2}, ~~
t > t_{eq}.
\eeq
The condition $t_{Th} \ll H^{-1}$ is obviously satisfied when $x_e\sim 1$ before the recombination. 
%when $x_e$ drops to $x_e \sim 10^{-5}$.

We now estimate  the interaction between dark matter in form of CCOs and photons.
As before, we assume that the cross section is proportional to the geometrical size, $4\pi R^2$
such that the relevant mean free time is 
\beq 
\label{DM}
t_{DM} =\frac{1}{4\pi R^2 n_{DM} c}
\simeq   \left(\frac{T}{1 eV}\right)^{-3}\left(\frac{4 \cdot 10^{9} B^{1/3}}{\Omega_{DM} h^2}\right)s.
\eeq
It is clear from this expression that for sufficiently large $B\gg 1$ the condition $t_{DM} \gg H^{-1}$ is obviously satisfied, and therefore, CCOs do indeed qualify as 
candidates for the role of cold dark matter. 
\subsection{ Generic feature of   CCO-- effectively weak interaction 
as geometrical factor }
The main message of  Section II  can be formulated as follows. 
We observed that there are no any contradictions with   observations   if DM   is  represented  by
   macroscopically large   composite compact objects made of ordinary matter
or even antimatter, as described above.  The main reason why this very counterintuitive  
concept   still does not contradict the observations has pure geometrical nature.
Indeed, the effective interaction which appears in all previous 
discussions is proportional to a factor,
$\epsilon\sim \sigma \cdot n_{DM}$ which can be represented in pure geometrical terms describing a composite objects
$\epsilon\sim{S}/{V}$. Indeed, a typical   scattering  cross section  off a large macroscopical object
is always proportional to its surface area,
  $\sigma\sim S\sim B^{2/3}$,     while  number density of heavy $DM$ particles,
  $n_{DM}\sim \rho_{DM}/M\sim V^{-1}\sim B^{-1}$ 
 is   proportional to the    inverse volume of the composite objects
 filled by quarks, $M\sim   V \sim B$.
Therefore,  for
macroscopically large composite objects this ratio  $\epsilon\sim \sigma \cdot n_{DM}
\sim{S}/{V}\sim B^{-1/3}\ll 1$ 
could be numerically very small if number of particles $B$ forming the object is very large.
This small geometrical factor  $\epsilon \ll 1$  can    
successfully  replace  the standard assumption on weak coupling interaction between visible and DM.
As it is known, this requirement 
  is   a crucial ingredient   of entire  idea  of DM as a substance which is collision-less 
 and weakly  interacting. Our remark here is: 
 the weakness of the interaction  can be achieved 
 by a new concept   of  compact composite objects with $\epsilon \ll 1$
 instead of introducing into the theory  some new,  not  yet discovered    weakly interacting massive particles (such as WIMPs).

\section{Thermodynamics} 
 The main goal of the previous section was to argue that CCO is qualified as a  cold dark matter 
 candidate. More precisely,  we argued that a new concept   of  compact composite objects does not contradict to  any observations and  does not spoil any standard requirements which are crucial for the structure formation. Therefore, naively one should not expect any differences in behavior 
between CCOs and   let us say,   the standard  WIMPs as far as structure formation is concern \footnote{Of course, there is distinctive feature between CCOs and other DM candidates 
 for other observables,  such as emission $511 ~KeV$ line, as we discussed in previous section.}.
 Nevertheless,
  we shall argue below that the ability of CCOs to interact efficiently with photons and hadrons,
 unlike fundamental point-like particles, is a new distinctive feature which might have 
 phenomenologically observable consequences relevant for the  structure formation at smaller scales. Hopefully, 
 it may even lead to the resolution of some problems   highlighted in the Introduction. 
 This new feature of CCOs provides a mechanism that, in principle, has ability to modify the structure formation in the central denser regions where the interaction of visible hadrons/photons with
  internal degrees of freedom of CCOs can reach thermal equilibrium. 

 Let us emphasize, we are not talking about thermal equilibrium between CCO
 as a whole object and visible matter.  There is no such equilibrium, as 
 we demonstrated previously. It should not be such an equilibrium  
 according to the    standard
 requirement  of the structure formation theory.  Rather, we are talking about thermal equilibrium
  between visible matter   and 
 {\it internal degrees of freedom} of a compact composite object.
   \subsection{Idea}
  The picture that we have in mind  is similar to 
the interaction of molecules of air with the internal 
excitations of a solid lattice, the phonons. Phonons can be, in good 
approximation, be described as   massless degrees of freedom 
in thermal equilibrium with surrounding environment   at the same room temperature $T$.
Obviously, this 
interaction is unable to keep the whole massive solid in thermal 
equilibrium with the molecules of the gas. The crucial point here is that
most of the mass of the solid is carried by the atoms that form the 
lattice, while the mass of vibrational excitations is orders of magnitude 
lighter.

Following this example, we consider a statistical ensemble of dense 
massive droplets  surrounded by a gas of much lighter 
hadrons. The typical mass of the droplet $M \simeq B m_N$, where $B$ is 
the baryon number carried by the droplet and $m_N$ is the mass of the 
nucleon. The baryon charge $B$ can  be very large. In particular, 
for the formation mechanism suggested in ref. \cite{Zhitnitsky:2002qa}, the typical $B\sim 10^{33}$.
However the discussions which follow have  much more generic applicability
and are  not limited by specific model considered in
 \cite{Zhitnitsky:2002qa}.  Therefore, in what follows we consider an arbitrary  
  compact composite object filled by  
 quarks and gluons (in color superconducting or any other phase) which can be treated   
as a CCO containing a gas of massless goldstone modes in thermal 
equilibrium with the surrounding gas of visible hadronic matter. Precise condition when such 
 thermal  equilibrium can be  maintained is formulated below.

The key point is as follows: the internal degrees of freedom of cold composite objects being in thermal  
equilibrium with hadrons store a new {\it hot} contribution  to 
the total   matter density. One can call such  dark matter
a {\it ``chameleon- like DM" }  because its properties strongly depend on environment surrounding it.
It has all features of ordinary cold DM in the very dilute  environment; 
it becomes hot   in the environment when   the ordinary visible  matter becomes very dense, and  thermal  equilibrium of internal degrees of freedom with
visible matter can be  maintained. To say it differently, the heat stored in CCOs can be transfered to visible matter if it is sufficiently 
dense, see precise qualitative relation below. Therefore, the DM in form of CCOs
 may change the standard picture of the dark matter  distribution at small scales when 
the visible matter  becomes very dense, such as it happens at the center of galaxies. 
\subsection{Thermodynamics of the internal degrees of freedom in the compact composite objects}

To be  precise, we say that the local thermal  
equilibrium of internal degrees of freedom of a compact object
 with visible hadrons is maintained at temperature $T^*$
\footnote{ do not confuse $T^*$  with the temperature describing the evolution of the entire universe
$T$.}   if the  number of baryons hard-hitting a single CCO  during the 
Hubble time greatly exceeds  the number of internal excitable degrees of freedom of this composite compact object.
This condition can be  expressed in a formal way  as follows,  
\beq 
\label{T}
(n_B \sigma v)\ \times  H^{-1} \gg \int_{V}  e^{-\frac{p}{T^*}}\cdot \frac{d^3p
d^3x}{(2\pi)^3} ,
\eeq
where combination $(n_B \sigma v)$ determines the total number of collisions of visible hadrons
 per second 
with a single compact object with a typical size $R$. In this formula we present $n_B$ as follows,
 $n_B\sim \xi{\overline{\rho}_B}/m_N
 \sim   \xi {\overline{\rho}_B}/{\rho_c}\cdot 10^{-5} h^2 cm^{-3}$;
 parameter $\xi$ in this expression is  $\xi\equiv \rho_B  /{\overline{\rho}_B}
\gg 1$ and it describes the excess  of the local baryon matter density $ \rho_B  $
(e.g. in galaxies) in comparison 
with   the  averaged density over entire universe
$ {\overline{\rho}_B}$. In formula (\ref{T}) we estimate the number of
internal massless degrees of freedom (such as phonos) of a CCO
assuming a simple Boltzmann distribution,
neglecting many complications related to the specific structure of the compact objects, such as 
presence of a condensate, spin of the Goldstone particles etc.  For numerical estimates we take 
$\sigma=4\pi R^2$ and $B\sim 4\pi R^3n_0/3 $ where $n_0\sim (108 MeV)^3 $ is nuclear density.
 Assuming the equilibrium as formulated above,
 we   estimate a  typical  velocity $v$ for the visible hadrons  as  $v/c \sim\sqrt{2T^*/m_N}$.
Having done all these   simplifications we arrive to the following numerical estimate
for the visible matter 
density  $\rho_B $   when  the local thermal  
equilibrium of internal degrees of freedom of the composite compact objects
can be  maintained  with visible hadrons in the dense environment  at temperature $T^*$,
\beq
\label{xi}
\xi \left(\frac{10 KeV}{T^*}\right)^{5/2} \gg 10^4 \left(\frac{B}{10^{33}}\right)^{1/3}, ~
\xi\equiv \rho_B  /{\overline{\rho}_B}.
\eeq
\subsection{Immediate   consequences} 
The obtained relation (\ref{xi}) is very suggestive,
  and it definitely deserves some additional comments.
First of all, if we interpret sign $\gg$ in eq.  (\ref{xi}) as factor $\sim ~10 $ or so,
the relation  (\ref{xi}) predicts that unusual features of the dark matter (assuming it is made
of  CCOs ) start to show up at densities where $\xi\sim 10^5$. Amazingly, this value of $\xi$
  precisely corresponds to  a typical matter density at   kpc scales  where the standard $N$ body (``cuspy") simulations apparently start to deviate   from  observational data. 
  Therefore, it is tempting to interpret this result as manifestation   of dark matter features
  (there existence of internal degrees of freedom hidden in CCOs) which 
  are not taking into account in the standard  
  N body simulations.  We shall 
  present few more arguments supporting this interpretation in the next section.

Another interesting scale  in this formula is $T^*\sim 10~ KeV$ when density $\xi$ approaches 
its typical galactic value $\xi\sim 10^5$. Why $T^*\sim 10~ KeV$ is so special? 
We would like to interpret this temperature $T^*\sim 10 ~KeV$ as a hot component of plasma 
which was the crucial ingredient  in 
the  recent fitting  of the  diffuse X-ray emission  from  the Galactic Center\cite{Muno:2004bs}.
This is  one of the problem highlighted in the Introduction.
According to ref. \cite{Muno:2004bs} the spectrum is consistent with   a two-temperature plasma with 
the hot component close to  $T\simeq 8 ~KeV$. According to analysis    \cite{Muno:2004bs}
the  hot component 
 is very difficult to understand within the standard picture. First of all, it 
would be too hot to be bound to the Galactic center. Authors of ref. \cite{Muno:2004bs} also remark that
the energy required to sustain such  plasma   corresponds to the entire kinetic energy of one supernova every 3000 yr, which is unreasonably high.    Finally, authors conclude with  the 
 following pessimistic note:
``We are left to conclude either that there is a significant shortcoming in our understanding of the mechanisms that heat the interstellar medium or that a population of faint   hard X-ray sources that is a factor of 10 more numerous".

 It is very tempting to interpret  this result as follows.  The missing sources of 
 the required heat   are stored in the form
of internal degrees of freedom of  CCOs as discussed above. This 
  interpretation is entirely based  on a  new concept of the dark matter   when  
excitable internal degrees of freedom   of CCOs and visible hadrons can be 
 in   the   local thermal   equilibrium. This only can happen 
  when the visible matter density is sufficiently large(\ref{xi}). The  
  interpretation  suggested here  is not  sensitive to the 
   specific details   of our model.  Instead,  the consequences described above are   very generic features 
of any dark matter substance if it is  represented by   the  large compact composite objects rather than by   
 some point like particles such as  WIMPS.
\section{ Other     Consequences}

Our new concept of the dark matter  is no doubt lead to many other 
important consequences for cosmology and astrophysics, which are not explored yet. We highlighted
 some of the problems (which can not be easily understood within the 
  conventional CCDM paradigm) in the Introduction. Those problems 
 may have some relation to  the  dark matter nature as  we   advocate in the present work.
 
$\bullet$ In particular,   we already  mentioned in Section IIC that excess of photons in MeV and GeV bands as well as 511 KeV line from the Galactic Center may have natural explanation if DM is consist of chunks 
of massive droplets.  We also mentioned in Section IIIC that unusual features (such as intensity and spectrum)  of the diffuse  $X$ ray emission   from the galactic center
 may also  have a natural explanation if DM is consist of chunks 
of massive droplets. 
 
Here we want to discuss some other
 issues related to {\it DM $\Leftrightarrow$ Baryons correlations } which  apparently  have been observed in many different instances, see few highlights in the Introduction. While 
   such correlations are quite natural outcome in our framework, see below, the same 
   correlations  look   very   mysterious and difficult to understand  within the 
  conventional Collisionless CDM paradigm. But first, we want to make few general remarks to
  emphasize  the fundamental  differences between our   concept of the dark matter
  and the conventional viewpoint when  the DM   is represented 
by  fundamental point-like particles like axion,  neutralino, 
or other WIMP candidates.

By definition, cold dark matter must have decoupled from the thermal 
plasma of hadrons, electrons and photons before recombination, so that it 
can gravitationally collapse in the characteristic triaxial virialized 
halos that drove the formation of large scale structures.  We argued in Section IID
that indeed the dark matter in form of CCOs does satisfy this requirement.
   This generic description trivially proves that there can exist a 
component of dark matter in internal degrees of freedom which is hot
in the sense that is in thermal equilibrium with hadrons, but it 
contributes zero pressure to the cosmic plasma because it is confined 
inside the massive and macroscopic cold droplets of condensed quarks. 
Evidently, such component is absent if  DM is represented 
by any fundamental point-like particles like axion,  neutralino, 
or other WIMP candidates.  
 
We have seen that supermassive   CCOs can be perfect candidates 
for the role of cosmological cold dark matter during the process of 
structure formation, as they never reach thermal equilibrium with the 
surrounding hadrons even if there is efficient transfer of energy   
  between the latter and the internal degrees of freedom of the 
former. This feature is quite unique and it is not shared by any other DM candidates.
 The gravitational collapse of the  
supermassive composite objects shall successfuly reproduce over large 
cosmological scales the characteristic features   of the web of  
hierarchichal CDM structures. 
On the other hand, within the overdense central regions of structure 
formation thermal equilibrium between hadrons and massless internal modes 
of composite objects can be reached and the interaction can modify the 
dynamics of the structure. According to condition (\ref{xi}), 
these regions roughly correspond to the sub-galaxies scales
where matter density is at least $\xi \simgt 10^5$ higher than 
the cosmological average. This condition roughly defines the 
inner core of galaxies where collissionless cold dark matter models fail 
to reproduce the high resolution features of the observed structures 
\cite{Primack:2002th}. 

$\bullet$ We anticipate that at this small scales when density is large,
$\xi \simgt 10^5$   the internal degrees of freedom of CCOs will heat up
the low entropy material, which would lead to additional pressure
in overdense regions. This interaction would result in   formation of a core rather than a cusp; 
it will also produce a shallower density profile; the centers of halos are expected
to be spherical (rather than triaxial) due to the same 
interactions.  At the virial radius of a typical galactic halo
where overdensity is much smaller, $\xi\ll 10^4$, the condition (\ref{xi}) is not satisfied
any more, and therefore, the usual, triaxial cold dark matter halo will result at these larger scales.
In many respects these results are very similar to the predictions  which follow from
strongly interacting DM models  \cite{Spergel:1999mh},  \cite{Dave:2000ar},  \cite{Ostriker:2003qj}.
 The microscopical nature of the interactions in these two cases, of course, is very different. 
 At the same time other  predictions (described in the next item) are not shared by
 strongly interacting DM models  \cite{Spergel:1999mh},  \cite{Dave:2000ar},  \cite{Ostriker:2003qj}.

$\bullet$ We   also anticipate 
  (due to the  interaction of visible matter with  dark matter   in our framework)     
  some correlations between DM and visible matter 
distributions. Intriguingly, a number of
different observations (see e.g. refs.
\cite{Sancisi:2003xt},\cite{Napolitano:2003xm},\cite{Salucci:2005mi}) apparently
do support such correlations.   In particular, in ref.\cite{Sancisi:2003xt} it is argued that
``there is a close correlation between rotation curve shape and  light distribution. For any feature in the luminosity profile there is a corresponding feature in the rotation curve and vice versa". 
Similarly, in ref. \cite{Napolitano:2003xm}  it was argued that the dark matter has triggered 
the evolution of both the stellar and hot gas components in galaxies. Very different analysis
\cite{Salucci:2005mi} reaches very similar conclusion ``that the galaxies  
... are uniquely successfully fitted by cored 
 haloes, with a core size comparable to the optical radius. 
 This suggests the existence of a well-defined scale length in 
 dark matter haloes, linked to the luminous matter, which
 is totally unexpected in the framework of CDM theory".
 Our remark here is as follows: all these observations  which apparently 
  are hinting on  DM $\Leftrightarrow$ Baryons correlations, 
 are  difficult to understand within the standard framework of collisionless CDM.
In contrast,  the same results   may have very natural explanation
  if dark matter is represented by droplets with excitable internal degrees of freedom
  when condition (\ref{xi}) is satisfied, in which case the DM and visible matter  distributions
    must be  obviously  correlated. 
   
  $\bullet$ The same interaction between DM and Baryons  can be also responsible 
  for the resolution of the angular momentum problem. Indeed, 
 such an  interaction can  prevent the visible gas from   overcooling,  which is considered
 to be the main cause of the angular momentum problem transfer\cite{Primack:2002th}.
 
 $\bullet$ Analysis of ref.\cite{Mishchenko:2003in} also finds
the  visible and dark matter correlations. This claim  
was supported by the fitting of the  dark matter vs visible matter distributions
   on the $\log -\log$ plot. The linear dependence between the two components was
   interpreted as an evidence of the thermal (or hydrodynamical)  equilibrium between the visible
   and DM components. While complete  understanding of this ``accidental correlation" 
   is obviously missing, the results of ref.\cite{Mishchenko:2003in} are quite encouraging from  
   the viewpoint advocated in this work when the internal degrees of freedom of droplets
   indeed  can be in the thermal equilibrium with visible matter in overdense regions.
   It is expected that the correlations would persist even at much larger scales after the  freeze-out
     (decoupling) takes place.
  
   It is   quite amazing that the effective temperature extracted  in
    ref.\cite{Mishchenko:2003in} turns out to be  of the same order of magnitude as  our estimate (\ref{xi})
    with $T^*\simeq 10 KeV \sim 10^8 K$. As we mentioned   in Section IIIC 
  precisely such temperature is required to fit the data on diffuse 
     $X$ ray emission from the galactic Center\cite{Muno:2004bs}.  
    Another intriguing result of the analysis  ref.\cite{Mishchenko:2003in}
    is the estimation of value  for  the dark matter  particle  mass ($300~ MeV$) 
    which is astonishingly close to the typical QCD scale. While we do not think that
    this value corresponds to any fundamental particle mass, we still expect that 
    such numerical value is  not  an ``accidental coincidence"  of the analysis but rather    is the  result 
  of the QCD dynamics 
    in the bulk of   the compact composite objects.      
     
 To conclude this section and  to avoid the  misunderstanding: 
 we are not claiming to have resolved all the problems highlighted
 in the Introduction (and discussed in this section). Rather, we wanted to present some arguments
 suggesting that many seemingly   unrelated problems
  might be in fact closely related if we accept the 
 new concept of DM nature which assumes that   dark matter particles to be represented
 by  compact composite objects formed  during the QCD phase transition.
  
 \section{ Conclusion. Future Directions.}
We conclude with few general remarks and few  suggestions for the future work. 
The main goal of this work was twofold. First, we argued that the new concept of the dark matter 
when it is represented in the form of macroscopically large compact composite objects
(with excitable internal degrees of freedom),
does not contradict to any current observations. Secondly,
this new concept has a potential for the natural explanations
of many phenomena which are difficult to explain   within the standard paradigm. 
Each piece of evidence  taken separately is   not convincing enough to 
  abandon the idea that DM is collisionless, non-interacting and  absolutely stable
  weakly interacting massive particle.  Nevertheless, it is very likely 
  that some of the problems discussed in this work would
  persist. In this case a new concept of DM nature must be developed, and this work 
  offers one of the  possibilities. 

 Unlike fundamental point-like cold matter particle,
 the  compact composite objects (CCOs) do interact with 
  visible hadrons  quite 
efficiently in the dense environment (\ref{xi})
by exciting the    internal   degrees of freedom confined to the bulk 
of CCOs. As we argued it does not    spoil  the coldness of CCOs. The 
internal modes represent a {\it hot} contribution to the total dark 
matter density that, nevertheless, contributes zero pressure to the cosmic 
fluid.

Over the large cosmological scales the interaction is mostly irrelevant
because matter, both visible and dark, is too sparsely distributed to keep
thermal equilibrium(\ref{xi}).
As a  consequence of it, the massive composite dark matter droplets shall 
produce a web of cosmic structures with the characteristic large scale 
hierarchichal features succesfully reproduced by any model of cold dark 
matter. 
On the other hand, within the overdense core regions of structure formation 
the visible hadrons can efficiently interact with composite dark matter objects, and reach 
thermal equilibrium. 

$\bullet$ This feature can change the dynamics of the structure 
in its central denser region satisfying eq. (\ref{xi}) 
as argued in Chapter IV. We presented a number of  qualitative
arguments supporting this claim. It is obvious:  only numerical simulations
can confirm or rule out the phenomena we predicted above.
 The crucial  new element 
of such simulations should be a modeling /incorporation
of  the interaction of visible matter with internal degrees of freedom
of the large droplets.    Therefore, we strongly advocate
to perform such an analysis to see whether the new phenomena anticipated in Chapter IV
indeed take place. 

$\bullet$ 
As we mentioned earlier there are  many other important phenomena in cosmology and astrophysics, which are not explored yet,     and which   may be also influenced by the new concept of DM
as composite objects. In particular,  as was
argued previously\cite{Zhitnitsky:2002qa},\cite{Oaknin:2003uv}
and reviewed in section IIB  there existence of  droplets made of anti -matter  
does not contradict to any current data, 
but rather may provide a natural explanation of the observed
ratio $\Omega_{DM}\sim \Omega_B$. At the same time the rare events of annihilation 
of such droplets could distort
the CMB anisotropies and polarization as first discussed in ref.\cite{Naselsky:2003zj}. 
Also: if we accept the  explanation of 511 $KeV$ line  and excess of photons
in $1 ~ GeV$ band as due to the annihilation of visible matter 
with anti matter droplets\cite{Oaknin:2004mn}, than there must be a correlation between 
511 $KeV$ flux distribution,  $1 ~ GeV$ photons and CMB anisotropies\cite{Naselsky}.

One more observational consequence: If the picture advocating in this work is 
indeed correct, then the DM isodensity contours should become more and more
circular (or at least follow more and more closely the baryon
distribution) at smaller radius. That's something one can hope to measure
with galaxy-galaxy lensing\footnote{I am thankful Ludvic Van Waerbeke for suggesting
such measurements  as a possible test of DM as compact composite objects. }. One should not
expect such a behavior if DM particles are ordinary WIMPs.

It could be many other observable consequences of this new concept
on nature of DM which are still to be explored.

{\it \underline{Latest Notes}} Some aspects of this framework (when dark matter is represented in the form of CCOs)
have been recently criticized by Cumberbatch, Silk and Starkman\cite{CSS}.
We refer to our  reply \cite{arz} on this criticism where it is  explicitly shown 
why the arguments presented in\cite{CSS} are incorrect.

 \section*{Acknowledgements}
 I am   thankful to Z. Berezhiani, V. Berezinsky,   V.Mukhanov, P. Naselsky and  G Senjanovic for useful discussions and critical remarks during The Summer Institute  on Particle Physics and Astrophysics at Gran Sasso, September, 2005,  where this work was presented. I am also thankful D. Chung, A. Dolgov, H. Richer, P. Steinhardt, M. Shaposhnikov, I. Tkachev and  L. Van Waerbeke for discussions   and   
  correspondence on the subject.  I am also thankful to David Oaknin who participated
    on the initial stage of  this project in 2003. 
  
  This work
 was supported, in part, by the Natural Sciences and Engineering
Research Council of Canada.

\end{document}